\documentclass[12pt]{article}
\usepackage[sumlimits,intlimits,namelimits]{amsmath}
\usepackage{amssymb}
\usepackage[english]{babel}
\usepackage{bbm}
\global\parskip 6pt

\newcommand{\be}{\begin{eqnarray}}
\newcommand{\ee}{\end{eqnarray}}
\newcommand{\non}{\nonumber}
\newcommand{\lb}{\label}
\def\>{\rangle}
\def\<{\langle}

\begin{document}
%
\begin{titlepage}
\vspace*{.5cm}
\begin{center}
{\Large{{\bf Thermodynamic Stability of Warped $AdS_{3}$ Black Holes}}} \\[.5ex]
\vspace{1cm}
Danny Birmingham\footnote{Email: dbirmingham@pacific.edu}\\
\vspace{.1cm}
{\em Department of Physics,\\
University of the Pacific,\\
Stockton, CA 95211\\
USA}\\
\vspace*{.5cm}
Susan Mokhtari\footnote{Email:
susan@science.csustan.edu}\\
\vspace{.1cm} {\em Department of Physics,\\
California State University Stanislaus,\\
Turlock, CA 95380,\\
USA} \\
\end{center}
\vspace*{1cm}
\begin{abstract}
\noindent {We study the thermodynamic stability of warped black
holes in three-dimensional topologically massive gravity.
The spacelike stretched black hole is parametrized by its mass and angular momentum.
We determine the local and global stability properties in the canonical and grand canonical
ensembles. The presence of a Hawking-Page type transition is established, and the critical temperature is
determined. The thermodynamic metric of Ruppeiner is computed, and the curvature is shown to diverge in the extremal limit.
The consequences of these results for the classical stability properties of warped
black holes are discussed within the context of the correlated stability conjecture.}
\\
\vspace*{.25cm}
\end{abstract}
\vspace*{.25cm}
\end{titlepage}

\section{Introduction}
Lower-dimensional models of gravity have provided a fruitful arena
for exploring various aspects of black holes. Of particular
importance are theories of three-dimensional anti-de Sitter gravity.
The presence of the BTZ black hole \cite{BTZ} as a solution to the
Einstein field equations has led to a wealth of important results on
the conformal properties of gravity. More recently, there has been
renewed interested in higher-derivative extensions of Einstein
gravity, through the addition of the gravitational Chern-Simons
terms, yielding topologically massive gravity \cite{Deser}-\cite{Carlip}, or other extensions such as new massive
gravity \cite{Townsend}.

For the case of topologically massive gravity (TMG), it is well know
that all solutions to the Einstein equations are also solutions of
TMG. The BTZ black hole therefore provides a useful example of a
black hole in this theory, and the classical stability against
linear perturbations was established in \cite{BMS}. In
\cite{Clement1}-\cite{Strom2}, a novel class of black hole solutions
to TMG was obtained. These have the attractive feature that they are
non-Einstein spaces, and thus they probe more details of the field
equations of TMG. Our primary concern here is to study the so-called
spacelike stretched black hole. This black hole is obtained as a
discrete quotient of a warped version of $AdS_{3}$, and the geometry
is parametrized by its mass, angular momentum, and a warp factor.
The thermodynamic variables of the black hole have been obtained,
and consistency with the first law of thermodynamics established
\cite{Clement1}-\cite{Strom2}.

However, various other aspects of its structure remain to be explored. In this paper, we determine the local and global
thermodynamic stability  properties of the warped black hole in both the grand canonical and canonical ensembles.
The local properties are determined by studying the Hessian of the entropy with respect to the mass and angular momentum, while the global properties
are uncovered by studying the Gibbs and Helmholtz free energies. We show that a Hawking-Page type transition is present in
both ensembles, and the critical temperature
is determined in terms of  the warp factor.

The plan of this paper is as follows. In section 2, we recall the metric and thermodynamic variables for the spacelike  stretched black hole.
In section 3, we show that the black hole is locally unstable in the grand canonical ensemble for all temperatures,
and globally stable above a critical temperature $T_{c}$.
In the canonical ensemble, we show that the black hole is locally unstable but globally stable for temperatures above $T_{c}$,
and vice versa for temperatures below $T_{c}$.  We also compute the thermodynamic curvature tensor of Ruppeiner \cite{Ruppeiner},
and show that it diverges in the extremal limit.
In section 4, we discuss the implications of these results within the context of the correlated stability conjecture of Gubser-Mitra \cite{Gubser1,
Gubser2}. We also record the local and global stability results for the BTZ black hole and show that it obeys the correlated stability conjecture.

\section{Warped Black Holes}

The action for topologically massive gravity is taken in the form \cite{Strom1}
 \be
S&=&\frac{1}{16\pi G}\int d^{3}x\sqrt{-g}\left(R+\frac{2}{l^2}\right)+\frac{1}{32\mu
\pi G}\int d^{3}x\sqrt{-g}\epsilon^{\lambda\mu\nu}\Gamma^{\rho}_{\lambda\sigma}
\left(\partial_{\mu}\Gamma^{\sigma}_{\nu\rho} +\frac{2}{3}\Gamma^{\sigma}_{\mu\tau}
\Gamma^{\tau}_{\nu\rho}\right),\lb{action}
\ee
where $\mu$ is the Chern-Simons coupling, and the parameter $l$ sets the scale of the cosmological constant of anti-de Sitter space,
$\Lambda = -1/l^{2}$. It the following, it will be convenient to use the parameter $\nu = \mu l/3$.

It is well known that any Einstein space, such as the BTZ black
hole,  is also a solution of the equations of motion for
topologically massive gravity, and its stability properties have
already been established \cite{BMS}. A novel class of warped black
hole solutions, which are non-Einstein, was discovered in
\cite{Clement1}-\cite{Strom2}. We will concentrate our attention on
the spacelike warped case, with line element given by \cite{Strom2}
\be ds^{2} =
\frac{l^{2}}{(\nu^{2}+3)}\left[-\cosh^{2}\sigma\;d\tau^{2} +
d\sigma^{2} + \frac{4\nu^{2}}{(\nu^{2} + 3)}(du +
\sinh\sigma\;d\tau)^{2}\right]. \label{warpedads} \ee For $\nu^{2} >
1$, the warp factor $\frac{4\nu^{2}}{(\nu^{2} + 3)}$ is greater than
one, and this spacetime is thus a spacelike stretched $AdS_{3}$
space. Black hole solutions asymptotic to spacelike stretched
$AdS_{3}$ space have been constructed as discrete quotients, leading
to the metric in Schwarzschild coordinates \cite{Strom2} \be
\frac{ds^{2}}{l^{2}} &=& dt^{2} + \frac{dr^{2}}{(\nu^{2} + 3)(r-r_{+})(r-r_{-})} + \left(2\nu r - \sqrt{(\nu^{2} + 3)r_{+}r_{-}}\right)dt\;d\theta\non\\
&+& \frac{r}{4}\left[3(\nu^{2} -1)r + (\nu^{2} + 3)(r_{+} + r_{-}) - 4\nu\sqrt{(\nu^{2} +3)r_{+}r_{-}}\right]d\theta^{2},
\label{warpedbh1}
\ee
where $r \in [0,\infty], t \in [-\infty, \infty]$ and $\theta \sim \theta + 2 \pi$. This spacetime represents a regular (spacelike stretched)
black hole when the warping parameter $\nu^{2} > 1$, and the parameters $r_{\pm}$ specify the location of the inner and outer horizon.

The spacetime can be written in standard ADM form as follows
\be
ds^{2} = - N(r)^{2}dt^{2} + l^{2} R(r)^{2}(d\theta + N^{\theta}(r)dt)^{2} + \frac{l^{4}dr^{2}}{4R(r)^{2}N(r)^{2}},
\label{warpadm}
\ee
where
\be
R(r)^{2} &=& \frac{r}{4}\left[3(\nu^{2} -1)r + (\nu^{2}+3)(r_{+} + r_{-}) - 4\nu\sqrt{(\nu^{2} + 3)r_{+}r_{-}}\right],\non\\
N(r)^{2} &=& \frac{l^{2}(\nu^{2} + 3)(r - r_{+})(r - r_{-})}{4R(r)^{2}},\non\\
N^{\theta}(r) &=& \frac{2\nu r - \sqrt{(\nu^{2} + 3)r_{+}r_{-}}}{2R(r)^{2}}.
\ee

The physical mass and angular momentum parameters can now be determined, with the results \cite{Clement2,Strom2}
\be
M &=& \frac{(\nu^{2} + 3)}{24G}\left(r_{+} + r_{-} -\frac{\sqrt{(\nu^{2} + 3)r_{+}r_{-}}}{\nu}\right),\non\\
J &=& \frac{\nu l (\nu^{2} + 3)}{96G} \left[ \left(r_{+} + r_{-} -\frac{\sqrt{(\nu^{2} + 3)r_{+}r_{-}}}{\nu}\right)^{2} -
\frac{(5\nu^{2} + 3)}{4\nu^{2}}(r_{+} - r_{-})^{2}\right].
\label{MJ}
\ee
The Hawking temperature and angular velocity take the form
\be
T &=& \frac{(\nu^{2} + 3)}{4\pi l} \frac{(r_{+} - r_{-})}{(2 \nu r_{+} - \sqrt{(\nu^{2} + 3)r_{+}r_{-}})},\non\\
\Omega  &=& \frac{2}{l(2 \nu r_{+} - \sqrt{(\nu^{2} + 3)r_{+}r_{-}})}.
\label{T}
\ee
In order to compute the entropy of the black hole, we must take into account the higher derivative terms in the action.
This leads to an entropy
\be
S = \frac{\pi l}{24 \nu G}\left[(9\nu^{2} + 3)r_{+} - (\nu^{2} + 3)r_{-} - 4\nu \sqrt{(\nu^{2} + 3)r_{+}r_{-}}\right].
\label{entropy1}
\ee

As highlighted in \cite{Strom2}, the goal is to have a conformal field theory interpretation of these black holes.
With this in mind, one can define left and right temperatures as follows
\be
T_{L} &=& \frac{(\nu^{2} + 3)}{8\pi l}\left(r_{+} + r_{-} -\frac{\sqrt{(\nu^{2} + 3)r_{+}r_{-}}}{\nu}\right),\non\\
T_{R} &=& \frac{(\nu^{2} + 3)}{8\pi l}(r_{+} - r_{-}),
\label{TLR}
\ee
which satisfy the relations
\be
\frac{1}{T} = \frac{4\pi \nu l}{(\nu^{2} + 3)}\left(\frac{T_{L} + T_{R}}{T_{R}}\right),\;\;
\frac{\Omega}{T} =\frac{1}{T_{R}l}.
\label{TOmega}
\ee
The central charges can be written in terms of the warp factor as
\be
c_{L} = \frac{4\nu l}{G(\nu^{2} + 3)},\;\; c_{R} = \frac{(5\nu^{2} + 3)l}{G\nu(\nu^{2} + 3)}.
\label{cLR}
\ee
As a result, the entropy can then be written in the suggestive form
\be
S = \frac{\pi^{2}l}{3}(c_{L}T_{L} + c_{R}T_{R}).
\label{entropy2}
\ee
The fact that the entropy can be reproduced from the Cardy formula in this way, led to the conjecture in \cite{Strom2}
that TMG for $\nu >1$ and with suitable asymptotically stretched $AdS_{3}$ boundary conditions is
holographically dual to a two-dimensional boundary conformal field theory with central charges given by (\ref{cLR}).

\section{Local and Global Thermodynamic Stability}

In order to determine the locally thermodynamic stability properties of a particular spacetime, we need to compute
the Hessian of the entropy with respect to the extensive variables $(M,J)$, see, for example, \cite{Gubser2,Callen}.
Using (\ref{MJ}) and (\ref{TLR}), the entropy
(\ref{entropy2}) can be written as
\be
S(M,J) = \sqrt{B\tilde{M}^{2} - C\tilde{J}} + D\tilde{M},
\label{entropy3}
\ee
where $\tilde{M} = Ml, \tilde{J} = Jl/G$, and
\be
B = \frac{4\pi^{2}(5\nu^{2} + 3)}{(\nu^{2} + 3)^{2}},\;\; C = \frac{2\pi^{2}(5\nu^{2} + 3)}{3\nu(\nu^{2} + 3)},\;\;
D = \frac{4\pi \nu}{\nu^{2} + 3}.
\label{BCD}
\ee
The Hessian matrix $\frac{\partial^{2}S}{\partial x^{i}\partial x^{j}}$, with $x^{i} = (\tilde{M},\tilde{J})$,
is denoted by $S^{\prime\prime}$ and takes the form
\be
S^{\prime\prime}= BC(B\tilde{M}^{2} - C\tilde{J})^{-3/2}\left(\begin{array}{cc}
-\tilde{J}&\frac{\tilde{M}}{2}\\
\frac{\tilde{M}}{2}& -\frac{C}{4B}
\end{array}\right).
\label{Sprime}
\ee
The determinant of the Hessian
\be
{\rm det}\;S^{\prime\prime}= -\frac{BC^{2}}{4}(B\tilde{M}^{2} -C\tilde{J})^{-2},
\label{detSprime}
\ee
is manifestly negative, with one positive and one negative eigenvalue.
Since local stability requires the Hessian to have only negative eigenvalues, we conclude that
the warped black hole is locally unstable in the grand canonical ensemble for all temperatures.

One can also obtain this result by starting with the mass formula $M(S,J)$ given by
\be
\tilde{M} = \left(\sqrt{BS^{2} + (BC - D^{2}C)\tilde{J}} - DS\right)/(B-D^{2}).
\ee
The Hessian of $\tilde{M}$ with respect to $(S,\tilde{J})$ is
\be
\tilde{M}^{\prime\prime}= BC[BS^{2} + (BC - D^{2}C)\tilde{J}]^{-3/2}\left(\begin{array}{cc}
\tilde{J}&-\frac{S}{2}\\
-\frac{S}{2}& -\frac{(BC-D^{2}C)}{4B}
\end{array}\right).
\label{Mprime}
\ee
The determinant of the Hessian is then
\be
{\rm det}\;\tilde{M}^{\prime\prime} = -\frac{1}{4}BC^{2}[BS^{2} + (BC - D^{2}C)\tilde{J}]^{-2}.
\ee
Local stability requires the Hessian of $\tilde{M}$ to be positive definite, which is clearly not the case.

Having established the local stability properties of the warped black hole, we turn our attention to
the issue of global stability. This can be resolved by considering the Gibbs free energy in the grand canonical ensemble, namely
\be
G = M - TS - \Omega J.
\label{Gibbs1}
\ee
Using the formulas (\ref{MJ})-(\ref{entropy1}), this can easily be expressed in terms of $r_{\pm}$. However, more work is required to determine $G$
as a function of $(T,\Omega)$. Using the relations (\ref{MJ}), and (\ref{TLR})-(\ref{entropy2}), we find
\be
G = -\frac{\pi^{2}l^{2}}{6G\nu}\left[T^{2} + \frac{2\nu}{\pi l}T - \frac{(\nu^{2} + 3)}{4\pi^{2}l^{2}}\right]\left(\frac{1}{\Omega l}\right),
\label{Gibbs2}
\ee
where the quadratic dependence on temperature is evident, and we note from (\ref{T}) that $\Omega >0$.
Solving the equation $G=0$ then yields the critical temperature
\be
T_{c} = -\frac{\nu}{\pi l} + \sqrt{\frac{5\nu^{2} + 3}{4 \pi^{2}l^{2}}}.
\label{Tc}
\ee
This leads to the result
\be
& &G < 0,\;\; {\rm for}\; T> T_{c},\\
& &G >0,\;\; {\rm for}\; T<T_{c}.
\label{Gglobal}
\ee
Thus, the warped black hole is dominant for $T> T_{c}$, while the warped thermal background is dominant for
$T < T_{c}$. This is reminiscent of the Hawking-Page transition which is present in the case of
the Schwarzschild anti-de Sitter black hole \cite{HP}. In the present case, we find that the warped black hole
is globally stable in the grand canonical ensemble only for temperatures greater than the critical temperature.

We can confirm this result by using the Smarr formula for the warped black hole, which takes the form \cite{Clement4}
\be
M = TS + 2\Omega J.
\label{Smarr1}
\ee
This allows us to re-write the Gibbs free energy as
\be
G = \Omega J.
\label{Gibbs3}
\ee
The sign of $G$ is therefore determined solely by the sign of the angular momentum $J$. Using the relation (\ref{TLR}),
the angular momentum can be expressed in terms of $(T_{L}, T_{R})$. Together with (\ref{TOmega}),
one can indeed show that
\be
J > 0 \leftrightarrow T < T_{c},\;\;J < 0 \leftrightarrow T > T_{c}.
\label{JT}
\ee
This establishes the
consistency of the global stability result obtained above.

It is also instructive to use the Gibbs free energy to determine the local stability. As shown in \cite{Monteiro},
local stability is equivalent to the statement that the Hessian of $(-G)$ with respect to $(T,\Omega l)$ is positive
definite. From (\ref{Gibbs2}), one can check that
\be
{\rm det} (-G^{\prime\prime}) = -\left(\frac{\pi l}{6G\nu}\right)^{2}\frac{(5 \nu^{2} + 3)}{(\Omega l)^{4}}.
\label{Gprime}
\ee
Since this is manifestly negative for all temperatures, it confirms the local instability result obtained previously.

Let us now turn our attention to the stability properties of the warped black hole in the canonical ensemble, see \cite{KimSon}.
The local stability is determined by
\be
\frac{\partial^{2}S}{\partial \tilde{M}^{2}} = -BC(B\tilde{M}^{2} - C\tilde{J})^{-3/2}\tilde{J}.
\label{canonical}
\ee
Hence, stability is determined by the sign of $J$, and from (\ref{JT}), we conclude that
the black hole is locally unstable for $T > T_{c}\; (J < 0)$ and locally stable for $T < T_{c}\; (J > 0)$.
The global stability is determined by the Helmholtz free energy
\be
F = M -TS.
\label{Helm}
\ee
Using (\ref{MJ})-(\ref{entropy1}), one obtains the surprising result that $F=2G$. Thus, the global stability properties in the canonical
ensemble are identical to the grand canonical ensemble, with the presence of a Hawking-Page transition
at the critical temperature $T_{c}$.
The proportionality between $F$ and $G$ is, however, more easily seen from the Smarr formula, which states that
$G = \Omega J$ and $F = 2\Omega J$.

Finally, it is useful to consider the metric on the space of thermodynamic variables introduced by Ruppeiner \cite{Ruppeiner}. The
Riemannian curvature of this metric has been used to probe
the phase structure of thermodynamic systems. The Ruppeiner metric for the warped black hole is defined by
\be
g_{ij}^{Ruppeiner} = -\frac{\partial^{2}S}{\partial x^{i}\partial x^{j}},
\label{Ruppeiner}
\ee
with coordinates $x^{i} = (\tilde{M},\tilde{J})$. The components of the metric can be read off from
(\ref{Sprime}) and,  as a result of the locally instability of the black hole, the metric has Lorentzian signature.
The Ricci scalar is given by
\be
R_{Ruppeiner} = -(B\tilde{M}^{2} - C\tilde{J})^{-1/2}.
\label{ScalarR}
\ee
Using the fact that
\be
B\tilde{M}^{2} - C\tilde{J} = \left(\frac{\pi l}{24G\nu}\right)^{2} (5\nu^{2} + 3)^{2}(r_{+} - r_{-})^{2},
\label{extreme}
\ee
we conclude that the  Ruppeiner curvature diverges in the extremal limit, as expected.

\section {The Correlated Stability Conjecture}
In \cite{Gubser1,Gubser2}, Gubser and Mitra established significant evidence for a correlation
between the classical and the local thermodynamic stability properties of black holes and black branes in anti-de Sitter space.
According to this conjecture, a black brane in anti-de Sitter space with a non-compact translational symmetry is classically stable if and only if
it is locally thermodynamically stable.
An analytic proof was attempted in \cite{Reall}. In the case of the warped $AdS_{3}$ black holes considered here, we note that the
coordinate $\theta \sim \theta + 2\pi$ could be identified with a translational coordinate.
Assuming that the correlated stability conjecture holds in this case, would allow us to conclude that the spacelike stretched black holes
are classically unstable. In order to verify this result, however, requires a study of the linear perturbations of the black hole.
Due to the reduced symmetry of the warped space, the perturbation equations have not yet been fully resolved into a convenient form
\cite{Anninos,Kim}.
However, the appropriate asymptotic boundary conditions have been determined \cite{Anninos}, and one can examine the behavior of a class of
highest weight solutions as a first step.
If it transpires that the warped black hole is actually classically stable, then one has a counterexample to the correlated stability conjecture.
Several counterexamples have already been discovered in \cite{Gubser3}.

As mentioned previously, the BTZ black hole is also a solution of the equations of motion for TMG, and the classical stability
against linear perturbations was established in \cite{BMS}.
It is of interest to determine whether the BTZ black hole obeys the correlated stability conjecture.
To see this, we consider the local stability properties as follows, \cite{Cai}-\cite{Aman}.
First, we recall that the entropy of the BTZ black hole can be written as \cite{Strom1}
\be
S = 2\pi\sqrt{\frac{c_{R}}{12}(Ml + J)} + 2\pi\sqrt{\frac{c_{L}}{12}(Ml - J)},
\label{BTZ-ent}
\ee
where the central charges are
\be
c_{L} = \frac{3l}{2G}(1 - \frac{1}{\mu l}),\;\; c_{R} = \frac{3l}{2G}(1 + \frac{1}{\mu l}).
\label{BTZ-c}
\ee
The Hessian of the entropy with respect to $(Ml,J)$ is given by
\be
S^{\prime\prime} = \left(\begin{array}{cc}
-x-y&-x+y\\
-x +y& -x -y
\end{array}\right),
\label{BTZ-Sprime}
\ee
where
\be
x = \frac{\pi}{2}\sqrt{\frac{c_{R}}{12}}(Ml +J)^{-3/2},\;\;y = \frac{\pi}{2}\sqrt{\frac{c_{L}}{12}}(Ml -J)^{-3/2}.
\label{xy}
\ee
Both eigenvalues of the Hessian are negative, with $\lambda = -2x, -2y$. Thus, the BTZ black hole
is locally stable in the grand canonical ensemble. Together with the classical stability result in \cite{BMS}, this confirms that the correlated
stability conjecture holds for the BTZ black hole in TMG.
Incidentally, the Smarr formula for the BTZ black hole \cite{Park}
\be
M = \frac{1}{2}TS + \Omega J,
\label{SmarrBTZ}
\ee
allows one to write the Gibbs free energy in the form
\be
G = -\frac{1}{2}TS.
\label{GibbsBTZ}
\ee
Since $G$ is manifestly negative, we conclude that the BTZ black hole is also globally stable in the grand canonical ensemble.

Finally, it should be pointed out that the warped black hole discussed here is also a solution of new massive gravity \cite{Clement3}.
The form of the metric is identical
to (\ref{warpadm}), except that the warping factor is determined by
a mass parameter $m$ instead of the Chern-Simons parameter $\mu$.

\section{Conclusions}

We have discussed the local and global thermodynamic stability properties of the spacelike stretched black hole
of topologically  massive gravity. In the grand canonical ensemble, it is locally unstable for all temperatures, and globally
unstable for temperatures below a critical value $T_{c}$ given by (\ref{Tc}). In the canonical ensemble, the black hole is
locally unstable and globally stable above for temperatures above $T_{c}$, and vice versa for temperatures below $T_{c}$.
The implications of these results for the classical stability of the black hole under linear perturbations remain to be checked.
According to the correlated stability conjecture, one would conclude that the black hole is classically  unstable due to the fact that it
is locally unstable in the grand canonical ensemble.
However, counterexamples to the correlated stability conjecture have been found, and therefore the task remains
to check classical stability. This study will be aided by the fact that the appropriate boundary conditions have already been determined \cite{Anninos},
although a resolution of the equations of motion in separable form presents a challenge \cite{Anninos,Kim}.
Stability properties for the null warped
solution of TMG have been discussed in \cite{Anninos2}.
\\

\noindent {\Large \bf  Acknowledgments}\\ \\
D.B and S.M are supported by the National Science Foundation under
grant PHY-0855133.

\end{document}